\documentclass[preprintnumbers,amsmath,11pt,amssymb,floatfix,superscriptaddress,nofootinbib]{revtex4}
\usepackage{graphicx}
\usepackage{epsfig}
\usepackage{bm}
\usepackage{amsfonts}

\input epsf

\begin{document}

\title{\Large Interacting Ricci Logarithmic Entropy-Corrected Holographic Dark Energy in Brans-Dicke Cosmology}

\author{ \textbf{Antonio Pasqua}}
\email{toto.pasqua@gmail.com} \affiliation{Department of Physics,
University of Trieste, Via Valerio, 2 34127 Trieste, Italy.}

\author{ \textbf{Iuliia Khomenko}}
\email{ju.khomenko@gmail.com} \affiliation{Heat-and-Power Engineering Department National Technical
University of Ukraine "Kyiv Politechnical Institute",� Kyiv, Ukraine.}

\begin{abstract}
\textbf{Abstract}: In the derivation of Holographic Dark Energy (HDE), the area law of the black
hole entropy assumes a crucial role. However, the entropy-area relation can be
modified including some quantum effects, motivated from the Loop Quantum
Gravity (LQG), string theory and black hole physics. In this paper, we study the cosmological
implications of the interacting logarithmic entropy-corrected HDE (LECHDE) model in
the framework of Brans-Dicke (BD) cosmology. As system's infrared (IR) cut-off, we choose the average radius of Ricci scalar curvature, i.e. $R^{-1/2}$. We obtain the Equation of State (EoS) parameter $\omega_D$, the deceleration parameter $q$ and the evolution of energy density parameter $\Omega'_D$ of our model in a non-flat universe. Moreover, we study the limiting cases corresponding to our model without corrections and to the Einstein's gravity.
\end{abstract}

\maketitle

\section{Introduction}
Cosmological and astrophysical data obtained, for example, with type Supernovae Ia  (SNeIa), the Cosmic Microwave Background (CMB) radiation anisotropies and Large Scale Structure (LSS) have provided strong evidences for  a phase of accelerated expansion of the universe \cite{1,1a,1b,1c}.\\
A large amount of the cosmic energy density is contained in the dark sectors, i.e. Dark
Energy (DE) and Dark Matter (DM) which represent, respectively, about the 73$\%$ and about the 23$\%$ of the cosmic energy density while the ordinary Baryonic Matter (BM) we are able to observe with our instruments contributes for the 4$\%$ of the total. Moreover, the contribution of radiation to the cosmic energy density can be considered practically negligible.\\
In relativistic cosmology, the cosmic acceleration we are able to observe can be described by a perfect fluid which pressure $p$ and energy density $\rho$ satisfy the following relation: $\rho + 3p < 0$. Such kind of fluid with negative pressure is named Dark Energy (DE). In other words, the relation  $\rho + 3p < 0$ implies that the EoS parameter $\omega_D =  p_D/\rho_D$ must obey the condition $\omega_D  <-1/3$, while from an observational point of view it is a daunting task to constrain its exact value. Several candidates have been studied in order to explain the nature of DE, including quintessence, cosmological constant $\Lambda$, tachyon, quintom,  K-essence, phantom energy, Chaplygin gas and modified gravity (see \cite{2,2a,2b} for more details).\\
The cosmological constant $\Lambda$ with EoS parameter $\omega = -1$ represents the earliest and simplest theoretical
candidate for DE. However, it is known that $\Lambda$ has two difficulties:
the fine-tuning and the cosmic coincidence problems \cite{copeland-2006}. The former asks why the vacuum energy density is so small (an order of $10^{123}$ smaller than what we observe) and the latter says why vacuum energy and DM are nearly equal today although they have evolved independently from different mass scales. Many attempts have been made till now by scientific community  in order to find a plausible explanation for the coincidence problem \cite{delcampo,delcampoa,delcampob,delcampoc,delcampod,delcampoe,delcampof,delcampog,delcampoh,delcampoi,delcampol}. \\
In literature, one of the most studied DE candidate is the Holographic DE (HDE) (which is motivated from the holographic principle) \cite{3,3a,3b,4,5,5a,5b,6}. It was shown by Cohen et al. \cite{7} that in Quantum Field Theory (QFT), the UV cut-off $\Lambda$ should be related to the IR cut-off $L$ due to limit set by forming a black hole. If the vacuum energy density caused by ultraviolet (UV) cut-off is given by $\rho_D = \Lambda^4$, then the total energy of size $L$ should not be greater than the mass of the system-size black hole, i.e.:
\begin{eqnarray}
E_D \leq E_{BH} \rightarrow L^3 \rho_D \leq M_p^2 L, \label{1}
\end{eqnarray}
where $M_p = \left( 8\pi G  \right)^{-1/2} \approx 10^{18}$GeV represents the reduced Planck mass (with $G$ representing the Newton's gravitational constant). If the largest possible cut-off $L$ is the one which
saturate this inequality, we obtain the energy density $\rho_D$ of HDE as follow:
\begin{eqnarray}
\rho_D = 3c^2 M_p^2 L^{-2}, \label{2}
\end{eqnarray}
where $c^2$ represents a dimensionless constant which value is still under debate. Following the recent work made by Guberina et al. \cite{8},  HDE based on the entropy bound can be derived in an alternative way. In the thermodynamics of the black hole \cite{9,9a}, a maximum entropy in a box of size $L$, known as Bekenstein-Hawking entropy bound, exists and it is given by $S_{BH} \approx M_p^2 L^2$, which scales as the area of the box $A \approx L^2$ rather than its volume
$V \approx L^3$. Moreover, for a macroscopic system with self-gravitation effects which can not be ignored,
the Bekenstein entropy bound $S_B$ is obtained multiplying the energy $E \approx \rho_DL^3$ and the linear size $L$ of the system. Requiring that the Bekenstein entropy bound is smaller than the Bekenstein-Hawking entropy (i.e. $S_B \leq S_{BH}$, which implies $EL \leq M_p^2 L^2$), it is possible to obtain the same result obtained from energy bound argument, i.e. $\rho_D \leq M_p^2L^{-2}$.\\
HDE was widely investigated in the literature in many different ways. In the work of Chen et al. \cite{10}, HDE was
used in order to drive inflation in the early evolutionary stages of universe. In the paper of Jamil et al. \cite{11}, the EoS parameter $\omega_D$ of HDE was studied considering a time-varying Newton's gravitational constant $G$, i.e. $G\left( t \right)$; moreover, it was shown that $\omega_D$ can be significantly modified in the low redshift limit.
HDE was also studied in other papers \cite{12,12a,12b,12c,12d,12e,12f,12g,13,13a,13b,13c}  with
different IR cut-offs, for example the particle horizon, the future event horizon, the Hubble horizon and the
recently proposed Granda-Oliveros (GO) cut-off. Similarly, correspondences between HDE and other
scalar field models of DE have been recently proposed\cite{14,14a,14b}, while in other papers, HDE was well studied in different
theories of modified gravity,  like $f\left(R\right)$ (where $R$ represents the Ricci scalar curvature), braneworld, DGP model, scalar-tensor gravity and Brans-Dicke (BD) \cite{15,15a,15b,15c,15d,15e,15f,15g,15h,15i,15l}. It was also demonstrated that HDE also fits well the cosmological data obtained from the SNeIa and CMB \cite{16,16a,16b,16c,16d,16e}.\\
It must be underlined that the black hole entropy $S$ assumes an important role in the derivation of HDE energy density. Indeed, we know that the derivation of HDE energy strongly depends on the entropy-area relation given, in Einstein's gravity, by $S \approx A \approx L^2$ (with $A$ indicating the area of the black hole horizon). However, the definition of the entropy-area relation can be modified taking into account quantum effects, motivated from the Loop Quantum Gravity (LQG).
These quantum corrections provided to the entropy-area relationship leads to the curvature
correction in the Einstein-Hilbert action and vice versa \cite{17,17a,17b}. The corrected entropy has the following form \cite{18,18a,18b,18c,18d}:
\begin{eqnarray}
S= \frac{A}{4G} + \widetilde{\alpha} \ln\left(\frac{A}{4G}\right) + \widetilde{\beta}, \label{3}
\end{eqnarray}
where $\widetilde{\alpha}$ and $\widetilde{\beta}$ are two dimensionless constants. The exact values of $\widetilde{\alpha}$ and $\widetilde{\beta}$ are not yet determined and they are still an open issue in quantum gravity. These considered corrections arise in the black hole entropy in LQG due to thermal and quantum
fluctuations \cite{19,19a,19b,19c,19d}. Moreover, considering Wald's approach to classical gravity and to string theory,
it is possible to find similar corrections to entropy \cite{20}. The entropy-area relation can be
generally expanded in a series of infinite terms, but the contribution given by extra terms can be considered practically negligible due to smallness of the value of the reduced Planck constant $\hbar$. Then, the most important order term in the expansion is the logarithmic one, as we have considered in this paper.
The logarithmic term also appears in a model of entropic cosmology, which is able to unify the early time
inflation and late-time cosmic acceleration of our universe \cite{21}. Considering the corrected
entropy-area relation and following the derivation of HDE, the energy density of HDE will result to be modified as well. For this reason, Wei \cite{22} recently proposed the energy density $\rho_D$ of the so-called logarithmic entropy-corrected HDE (LECHDE)
in the following form:
\begin{eqnarray}
    \rho_D=3c^2 M_p^2L^{-2}+ \alpha L^{-4} \log \left( M_p^2L^2 \right) + \beta L^{-4},\label{4}
\end{eqnarray}
where $\alpha$ and $\beta$ are dimensionless constants. In the limiting case corresponding to $\alpha= \beta = 0$, Eq. (\ref{4}) yields the well-known HDE density. The second term and third term in Eq. (\ref{4})
 are comparable to the first one only when $L$ assumes a very small value, then the corrections given by the second term and third term have a physical meaning only at early evolutionary stages of our universe. When universe becomes large, LECHDE reduces to the ordinary HDE.
Since HDE density belongs to a dynamical cosmological constant, we need a dynamical
frame to accommodate it instead of General Relativity. Moreover, considering as cut-off $L = H^{-1}$, it is not able to
determine the EoS parameter $\omega_D$ in the General Relativity framework. Futhermore, the BD scalar
field speeds up the expansion rate of a dust matter dominated era (i.e. reduces deceleration),
while slows down the expansion rate of cosmological constant era (i.e. reduces acceleration).
Taking into account all the consideration mentioned above, the investigation of HDE models in the framework
of BD theory is well motivated. For this reason, HDE has been widely studied in the framework of BD gravity \cite{23,23a,24,24a,24b,24c,24d,24e,25}. In these papers, many dynamical features of HDE
have been explored in the flat/non-flat Friedmann-Lemaitre-Robertson-Walker (FLRW) background, e.g. the cosmic coincidence problem, the quintom behavior, the phantom crossing
at the present time, the effective EoS parameter $\omega_{eff}$ and the the deceleration parameter $q$.\\
In this paper, we propose the R-LECHDE model which is obtained using as IR cut-off for the LECHDE the average radius of Ricci scalar curvature, i.e. $L = R^{-1/2}$. For a non-flat universe, the Ricci scalar curvature $R$ is given by:
\begin{equation}
 R=6\left(\dot{H}+2H^2+\frac{k}{a^2}\right),\label{5}
\end{equation}
where $H=\dot{a}/a $ represents the Hubble parameter (the overdot represents the derivative with respect to the cosmic time $t$), $\dot{H}$ is the derivative of the Hubble parameter with
respect to the cosmic time $t$, $a$ is a dimensionless scale factor (which is function of the
cosmic time $t$) and $k$ is the curvature parameter which can assume the values -1, 0, +1 which
yield, respectively, a closed, a flat or an open FLRW universe. The curvature paramater $k$
has dimension of $length^{-2}$ and describes the spatial geometry of space-time.\\
This work can be considered as extension of the work ok Feng \cite{fengBD} who studied the Ricci Dark Energy in Brans-Dicke cosmology: we now consider the contribution of the logarithmic correction to the entropy.\\
The average radius of the Ricci scalar curvature was proposed for the first time as IR
cut-off in one recent paper of Gao et al. \cite{gao}. It was found that this kind of model works well when it is fitted with observational data. Furthermore,  it can also be helpful in the understanding of the coincidence problem and
the presence of the event horizon is not presumed in this model, so it is possible to avoid
the casuality problem.\\
After the recently proposed work of Cai, Hu and Zhang \cite{cai}, where the casual entropy bound in
the holographic framework is studied, the Ricci model gained an appropriate reason for which it could
be motivated, giving an appropriate physical motivation for the Holographic Ricci DE
(RDE).\\
RDE has been widely studied in literature: in fact, we can find works where RDE was used for reconstruction of scalar fields \cite{mio2}, statefinder diagnostic \cite{feng3}, reconstruction of $f\left(R\right)$ \cite{feng1}, quintom \cite{feng2}, contributions of viscosity  \cite{feng4} and related observational constraints \cite{xu}.\\
This paper is outlined as follow. In Section 2, we study the R-LECHDE in the framework of BD
theory for non-flat universe. In Section 3, we discuss some of the features of this model including
the EoS parameter $\omega_D$, the evolution of dimensionless energy density $\Omega'_D$ and the deceleration parameter $q$ in the presence of interaction between DE and DM. Moreover, we calculate the limiting cases corresponding to $\alpha = \beta =0$ (i.e. $\gamma_c =1$), which means no corrections to the energy density, and to the Einstein's gravity. Finally, Section 4 is devoted to Conclusions.

\section{Brief Overview of Brans-Dicke Cosmology}
Within the framework of the standard Friedmann-Lematire-Robertson-Walker (FLRW) cosmology, the line element for non-flat universe is given by:
\begin{eqnarray}
    ds^2=-dt^2+a^2\left(t\right)\left[\frac{dr^2}{1-kr^2} +r^2 \left(d\theta ^2 + \sin^2 \theta d\varphi ^2\right) \right],\label{6}
\end{eqnarray}
where $t$ represents the cosmic time, $r$ is referred to the radial component and $\left(\theta , \varphi \right)$ are the two angular coordinates.\\
The BD action $S$ is given by:
\begin{eqnarray}
S=\int d^4x\sqrt{g}\left( -\varphi R+\frac{\omega}{\varphi} g^{\mu \nu}\partial_{\mu}\varphi \partial_{\nu}\varphi + L_M    \right). \label{7}
\end{eqnarray}
Using the definition:
\begin{eqnarray}
\varphi = \frac{\phi^2}{8\omega}, \label{8}
\end{eqnarray}
we recover the action $S$ in the canonical form \cite{26,26-1}:
\begin{eqnarray}
	S=\int d^4x\sqrt{g}\left( -\frac{1}{8\omega}\phi^2R+\frac{1}{2}g^{\mu \nu}\partial_{\mu}\phi \partial_{\nu}\phi+L_M    \right), \label{9}
\end{eqnarray}
where $g$, $\omega$, $\phi$, $R$ and $L_M$ are, respectively, the determinant of the tensor metric $g^{\mu \nu}$, the BD parameter, the BD scalar field, the Ricci scalar curvature and the Lagrangian of the matter. The variation of $S$ with respect to the FLRW metric given in Eq. (\ref{6}) yields the following Friedmann equations in the framework of BD theory:
\begin{eqnarray}
\frac{3}{4\omega}\phi^2\left(H^2+\frac{k}{a^2}\right)-\frac{\dot{\phi}^2}{2}+\frac{3}{2\omega}H\dot{\phi}\phi    &=& \rho_D + \rho_M, \label{10} \\
-\frac{\phi^2}{4\omega}\left(\frac{2\ddot{a}}{a}+H^2 +\frac{k}{a^2}\right)-\frac{H\dot{\phi}\phi}{\omega}-\frac{\ddot{\phi}\phi}{2\omega}-\frac{\dot{\phi}^2}{2}\left(1+\frac{1}{\omega} \right) &=& p_D, \label{11}\\
	\ddot{\phi}+3H\dot{\phi}- \frac{3}{2\omega}\left(\frac{\ddot{a}}{a}+H^2+\frac{k}{a^2}  \right)\phi  &=& 0. \label{12}
\end{eqnarray}
$\rho_D$ and $p_D$ represent, respectively, the energy density and pressure of DE. Instead, $\rho_m$ represents the energy density of DM, which is in this paper considered pressureless ($p_m = 0$). \\
Eqs. (\ref{10}), (\ref{11}) and (\ref{12}) form a system of equations which is not closed, then we have the freedom to choose another one. We now assume that the BD field $\phi$ can be described by a power-law of the scale factor $a$, i.e. $\phi \propto a^{n}$. In principle, there are not compelling reasons for this choice of $\phi$. However, it has been shown that for small values of $n$, it leads to consistent results \cite{28,29}. A case of particular interest is obtained when $n$ assumes small values, while $\omega$  high ones, so that $n\omega \approx 1$. This is interesting since local astronomical experiments set a very high lower bound on the value of $\omega$; in particular, thanks to the Cassini experiment, we know that $\omega > 10^4$ \cite{30,30-1}.\\
Taking the first and the second derivative with respect to the cosmic time $t$ of $\phi$, we get:
\begin{eqnarray}
	\dot{\phi} &=& n a^{n -1}\dot{a}= n H \phi, \label{13} \\
	\ddot{\phi} &=&  nH\dot{\phi}  +  n \phi\dot{H} = n ^2 H^2 \phi + n \phi \dot{H}. \label{14}
\end{eqnarray}
Next Section is devoted to the derivation of some important quantities, i.e the EoS parameter $\omega_D$, the evolution of energy density parameter $\Omega'_D$ and the deceleration parameter $q$.

\section{Interacting R-LECHDE Model}
We know that in BD cosmology the scalar field plays the role of the Newton's constant $G$ (i.e. $\left(\phi^2 \propto 1/G\right)$), so it becomes natural to express the energy density of R-LECHDE in BD cosmology in the following way:
\begin{eqnarray}
\rho_D = \frac{3c^2\phi^2}{4\omega L^2}   + \alpha L^{-4}\ln \left(\frac{\phi^2 L^2}{4\omega}\right) + \beta L^{-4}, \label{15}
\end{eqnarray}
which can be rewritten as:
\begin{eqnarray}
\rho_D = \frac{3c^2\phi^2}{4\omega L^2} \gamma_c , \label{16}
\end{eqnarray}
where:
\begin{eqnarray}
\gamma_c = 1   + \frac{4\omega \alpha }{3c^2\phi^2 L^2}\ln \left(\frac{\phi^2 L^2}{4\omega}\right) + \frac{4\omega \beta }{3c^2\phi^2 L^2} .  \label{17}
\end{eqnarray}
In the limiting case corresponding to $\alpha = \beta = 0$ ($\gamma_c =1$), Eq. (\ref{15}) reduces to:
\begin{eqnarray}
\rho_D = \frac{3c^2\phi^2}{4\omega L^2} , \label{15-1}
\end{eqnarray}
which is the the well-known HDE density in the BD cosmology \cite{sceichi}.\\
Substituting $L$ with $R^{-1/2}$ in Eqs. (\ref{15}), (\ref{16}) and (\ref{17}), we obtain:
\begin{eqnarray}
\rho_D  &=& \frac{3c^2\phi^2}{4\omega }R   + \alpha R^{2}\ln \left(\frac{\phi^2}{4\omega R}\right) + \beta R^{2} ,  \label{18} \\
\rho_D &=& \frac{3c^2\phi^2 R}{4\omega} \gamma_c,  \label{19}\\
\gamma_c &=& 1   + \frac{4\omega \alpha R}{3c^2\phi^2}\ln \left(\frac{\phi^2}{4\omega R}\right) + \frac{4\omega \beta R }{3c^2\phi^2 }.  \label{20}
\end{eqnarray}
In Eqs. (\ref{15}), (\ref{16}), (\ref{17}), (\ref{18}), (\ref{19}) and (\ref{20}), we have that  $\phi^2 = \omega/2\pi G_{eff}$ and $G_{eff}$ is the effective gravitational constant. In the limiting case of Einstein's gravity where $G_{eff}\rightarrow G$, we have  $\phi^2 = 4\omega M_p^2$ and the LECHDE energy density in Einstein's gravity is restored.\\
The critical energy density $\rho_{cr}$ and the energy density of the curvature $\rho_k$ are defined, respectively, as:
\begin{eqnarray}
\rho_{cr} &=& \frac{3\phi^2 H^2}{4\omega}, \label{21}   \\
 \rho_k &=& \frac{3k\phi^2}{4\omega a^2}.  \label{22}
\end{eqnarray}
Moreover, the fractional energy densities for DM, curvature and DE are defined, respectively, as follow:
\begin{eqnarray}
\Omega_m &=&   \frac{\rho_m}{\rho_{cr}}   = \frac{4\omega \rho_m}{3\phi^2 H^2},  \label{23}\\
\Omega_k &=& \frac{\rho_k}{\rho_{cr}}   = \frac{k}{a^2H^2},  \label{24}\\
\Omega_D &=&   \frac{\rho_D}{\rho_{cr}}   = \frac{c^2\gamma_c}{L^2H^2}.   \label{25}
\end{eqnarray}
Combining Eqs. (\ref{13}), (\ref{21}) and (\ref{22}) with the Friedmann equation given in Eq. (\ref{10}), we obtain:
\begin{eqnarray}
\rho_{cr} + \rho_k = \rho_m + \rho_D + \rho_{\phi}, \label{26}
\end{eqnarray}
where:
\begin{eqnarray}
\rho_{\phi} = \frac{1}{2}nH^2\phi^2\left( n - \frac{3}{\omega}   \right).\label{27}
\end{eqnarray}
Dividing Eq. (\ref{26}) by the critical energy density $\rho_{cr}$, we have that Eq. (\ref{26}) can be rewritten as:
\begin{eqnarray}
1 + \Omega_k = \Omega_m + \Omega_D + \Omega_{\phi},\label{28}
\end{eqnarray}
where:
\begin{eqnarray}
 \Omega_{\phi} = 2n \left( \frac{n\omega}{3}  -1 \right).\label{29}
\end{eqnarray}
In order to preserve the local energy-momentum conservation law, i.e. $\nabla_{\mu}T^{\mu \nu}=0$, the total energy density $\rho_{tot}= \rho_D + \rho_m$ must satisfy the following continuity equation:
\begin{eqnarray}
    \dot{\rho}_{tot}+3H\left( 1+\omega_{tot} \right)\rho_{tot}=0,\label{30}
\end{eqnarray}
where $\omega_{tot} \equiv p_{tot}/\rho_{tot}$ represents the total EoS parameter. Since we are considering interaction between DE and DM, the two energy densities for DE and DM $\rho_D$ and $\rho_m$ are preserved separately and the relative equations of conservation become:
\begin{eqnarray}
    \dot{\rho}_D&+&3H\rho_D\left(1+\omega_D\right)=-Q, \label{31} \\
\dot{\rho}_m&+&3H\rho_m=Q, \label{32}
\end{eqnarray}
where $Q$ is an interaction term which can be an arbitrary function of cosmological parameters, like the Hubble parameter $H$ and energy densities $\rho_m$ and $\rho_D$. We decide to use one of the most used expression for $Q$, given by:
\begin{eqnarray}
    Q = 3b^2H(\rho_m + \rho_D),\label{33}
\end{eqnarray}
where $b^2$ is a coupling parameter between DE and DM \cite{q1,q1-1,q1-2,q1-3,q1-4,q1-5,q1-6,q1-7,q1-8,q1-9,q1-10}. If $b^2$ is positive, we have transition from DE to DM, instead if $b^2$ assume negative values we have transition from DM to DE. The case with $b^2=0$ (i.e $Q=0$) represents the non-interacting FLRW model, instead $b^2=1$ yields a complete transfer from DE to DM. It was recently reported that this interaction is observed in the Abell cluster A586 showing a transition of DE into DM and vice versa \cite{q2,q2-2}. However, the strength of this interaction is not clearly identified \cite{q3}.\\
Observations of CMB and galactic clusters show that $b^2<0.025$, i.e. $b^2$ is a small positive constant \cite{q4}. We must also note that the ideal interaction term must be motivated from the theory of quantum gravity, otherwise we rely on dimensional basis for choosing an interaction term $Q$. We must emphasize that more general phenomenological terms can be used instead of the one we choose in this work since the nature of DE and DM is still not well-understood. \\
We now want to derive the expression of the EoS parameter $\omega_D$ for our model.\\
As stated in Introduction, the Ricci scalar curvature $R$ is given by:
\begin{eqnarray}
R= 6\left( \dot{H} + 2H^2 + \frac{k}{a^2}  \right). \label{34}
\end{eqnarray}
We derive from Eq. (\ref{10}), using the definition of $\phi$ and Eq. (\ref{13}), that:
\begin{eqnarray}
H^2 + \frac{k}{a^2} = \frac{4\omega}{3\phi^2}\left( \rho_D + \rho_m \right) + 2nH^2\left(\frac{n\omega}{3} -1 \right). \label{35}
\end{eqnarray}
Then, the Ricci scalar curvature $R$ can be also written as follow:
\begin{eqnarray}
R= 6\left[ \dot{H} + H^2 +  \frac{4\omega}{3\phi^2}\left( \rho_D + \rho_m \right) + 2nH^2\left(\frac{n\omega}{3} -1 \right)  \right]. \label{36}
\end{eqnarray}
We now want to derive an expression for the quantity $\dot{H} + H^2$.
Differentiating Eq. (\ref{35}) with respect to the cosmic time $t$ and using the continuity equations given in Eqs. (\ref{31}) and (\ref{32}), we derive:
\begin{eqnarray}
\dot{H}-\frac{k}{a^2} = 2n\dot{H}\left(\frac{n\omega}{3} -1 \right) -\frac{2\omega}{\phi^2}\left[\rho_D \left(1 + \omega_D \right) + \rho_m  \right] - \frac{4\omega n}{3\phi^2}\left( \rho_D + \rho_m  \right). \label{37}
\end{eqnarray}
Adding Eqs. (\ref{35}) and (\ref{37}), we obtain:
\begin{eqnarray}
\dot{H} + H^2 = \frac{\frac{4\omega}{3\phi^2}\left[ \rho_D\left( \frac{1}{2}+n + \frac{3\omega_D}{2}\right) + \rho_m\left( \frac{1}{2}+n\right) \right]}{2n\left(\frac{n\omega}{3} -1 \right) - 1}. \label{38}
\end{eqnarray}
Inserting Eq. (\ref{38}) in the expression of the Ricci scalar curvature given in Eq. (\ref{36}), we obtain:
\begin{eqnarray}
R =6 \left[\frac{\frac{4\omega}{3\phi^2}\left[ \rho_D\left( \frac{1}{2}+n + \frac{3\omega_D}{2}\right) + \rho_m\left( \frac{1}{2}+n\right) \right]}{2n\left(\frac{n\omega}{3} -1 \right) - 1} +  \right. \nonumber \\
 \left. + \frac{4\omega}{3\phi^2}\left( \rho_D + \rho_m \right)   + 2nH^2\left(\frac{n\omega}{3} -1 \right)  \right]. \label{39}
\end{eqnarray}
We can now easily derive the expression of $\omega_D$ from Eq. (\ref{39}):
\begin{eqnarray}
\omega_D &=&\left[ \frac{R}{36\omega \rho_D}      -\frac{n H^2\left(  n\omega -3 \right)}{9\omega \rho_D}          \right] \phi^2\left(2n^2\omega -6n -3 \right) \nonumber \\
&&- \frac{\left( 4n^2\omega -6n-3 \right)}{9}\left(\frac{\rho_D + \rho_m}{\rho_D}\right). \label{40}
\end{eqnarray}
Using Eqs. (\ref{27}) and (\ref{28}), we can rewrite Eq. (\ref{40}) as:
\begin{eqnarray}
\omega_D &=&\left[ \frac{R\phi^2}{36\omega \rho_D} -   \frac{2\rho_{\phi}}{9 \rho_D}   \right]  \left(2n^2\omega -6n -3 \right) \nonumber \\
&& - \frac{\left( 4n^2\omega -6n-3 \right)}{9}\left(\frac{1+\Omega_k - \Omega_{\phi}}{\Omega_D}\right) . \label{41}
\end{eqnarray}
In order to obtain the final expression of $\omega_D$, we must now substitute the expression of $\rho_D$ given in Eq. (\ref{18}) in Eq. (\ref{41}), obtaining:
\begin{eqnarray}
\omega_D &=&\left[ \frac{1}{27c^2 \gamma_c}\left(1 -\frac{8\rho_{\phi}}{\phi^2R }\right) \right]\left( 2n\omega^2 -6n -3 \right) \nonumber \\
 &&  - \frac{1}{9}\left( 4n^2\omega -6n-3 \right)\left(\frac{1+\Omega_k - \Omega_{\phi}}{\Omega_D}\right). \label{42}
\end{eqnarray}
We now want to derive an expression for the evolution of energy density parameter $\Omega'_D$.\\
The derivative with respect to the cosmic time $t$ of the energy density $\rho_D$ given in Eq. (\ref{18}) can be written, using the conservation equations given in Eqs. (\ref{33}) and (\ref{34}), as:
\begin{eqnarray}
    \dot{\rho}_D=3H\left[ -\rho_D- b^2\left(\rho_D + \rho_m\right) - \rho_D \omega_D   \right].\label{43}
\end{eqnarray}
Inserting the expression of $\omega_D$ given in Eq. (\ref{40}), Eq. (\ref{43}) yields:
\begin{eqnarray}
    \dot{\rho}_D&=&3H\left \{ -\rho_D +  \phi^2\left(2n^2\omega -6n -3 \right) \left[ \frac{nH^2\left(n\omega -3\right)}{9\omega} - \frac{R}{36 \omega}  \right] \nonumber \right. \\
   && \left.  + \left[ \frac{\left( 4n^2\omega -6n-3 \right)}{9} -b^2\right]\left(\rho_D + \rho_m \right)   \right\}. \label{44}
\end{eqnarray}
Dividing Eq. (\ref{44}) by the critical density $\rho_{cr}$ and using Eq. (\ref{28}), we obtain:
\begin{eqnarray}
 \frac{\dot{\rho}_D}{\rho_{cr}}  &=& \dot{\Omega}_D + 2\frac{\dot{H}}{H}\Omega_D + 2nH \Omega_D =  \nonumber \\
 &&  = 3H\left \{ -\Omega_D +\left(2n^2\omega -6n -3 \right)\left( \frac{2}{9}\Omega_{\phi}-\frac{1}{3} \frac{R}{9H^2} \right)+ \right. \nonumber \\
   && \left. \left[ \frac{\left( 4n^2\omega -6n-3 \right)}{9} -b^2\right]\left(1 + \Omega_k - \Omega_{\phi} \right)   \right \}. \label{45}
\end{eqnarray}
From the definition of the Ricci scalar curvature $R$, we derive that the term $\frac{R}{9H^2}$ is equivalent to:
\begin{eqnarray}
    \frac{R}{9H^2}=\frac{2}{3}\left( \frac{\dot{H}}{H^2} +2 + \Omega_k \right).\label{46}
\end{eqnarray}
Inserting Eq. (\ref{46}) in Eq. (\ref{45}), we have:
\begin{eqnarray}
 \dot{\Omega}_D &=& - 2\frac{\dot{H}}{H}\left[ \Omega_D + \frac{\left( 2n^2\omega -6n -3   \right)}{3}\right]+     \nonumber \\
&&  +3H\left \{ -\Omega_D\left(  1+\frac{2n}{3}  \right)  + \frac{2 \left(2n^2\omega -6n -3 \right)}{9} \left(\Omega_{\phi} -2 - \Omega_k  \right) + \right. \nonumber \\
&&\left.   \left[\frac{\left( 4n^2\omega -6n-3 \right)}{9}-b^2\right]\left(1 + \Omega_k - \Omega_{\phi} \right)   \right \}. \label{47}
\end{eqnarray}
Since  $\Omega_D'=\frac{d\Omega_D}{dx}= \frac{1}{H}\dot{\Omega}_D$ (where $x=\ln a$), we derive from Eq. (\ref{47}):
\begin{eqnarray}
 H\Omega'_D&=& - 2H'\left[ \Omega_D + \frac{\left( 2n^2\omega -6n -3   \right)}{3}\right]  \nonumber \\
 &&+3H\left\{ -\Omega_D\left(  1+\frac{2n}{3}  \right) + \frac{2 \left(2n^2\omega -6n -3 \right)}{9} \left(\Omega_{\phi} -2 -\Omega_k  \right)  \right. \nonumber \\
 &&\left.   + \left[\frac{\left( 4n^2\omega -6n-3 \right)}{9}-b^2\right]\left(1 + \Omega_k - \Omega_{\phi} \right) \right\}. \label{48}
\end{eqnarray}
Eq. (\ref{48}) yields:
\begin{eqnarray}
 \Omega'_D&=& - \frac{2}{H}\left[ \Omega_D + \frac{\left( 2n^2\omega -6n -3   \right) }{3}  \right]    \nonumber \\
  &&+3\left\{ -\Omega_D\left(  1+\frac{2n}{3}  \right) +  \frac{2 \left(2n^2\omega -6n -3 \right)}{9} \left(\Omega_{\phi} -2 -\Omega_k  \right)   \nonumber \right. \\
  && \left. + \left[\frac{\left( 4n^2\omega -6n-3  \right)}{9}-b^2\right]\left(1 + \Omega_k - \Omega_{\phi} \right)\right\}. \label{49}
\end{eqnarray}
which represents the expression of $\Omega'_D$ we wanted to derive.\\
In Eq. (\ref{49}), we used the fact that:
\begin{eqnarray}
   H' =  \frac{a'}{a}=1.  \label{50}
\end{eqnarray}
For completeness, we now derive the expression of the deceleration parameter $q$, defined as:
\begin{eqnarray}
q = - \frac{\ddot{a}a}{\dot{a}^2} = -\frac{\ddot{a}}{aH^2} = -1 - \frac{\dot{H}}{H^2}. \label{51}
\end{eqnarray}
$q$, combined with the Hubble parameter $H$ and the dimensionless density parameters given in Eqs. (\ref{23}), (\ref{24}) and (\ref{25}), form a set of very useful parameters for the description of the astrophysical observations. Dividing Eq. (\ref{11}) by $H^2$ and  using Eqs. (\ref{13}), (\ref{14}), (\ref{15}) and (\ref{25}), it is possible to write the deceleration parameter as:
\begin{eqnarray}
q = \frac{1}{2\left( n+1\right)}\left[ \left( 2n+1\right)^2 + 2n\left( n\omega - 1\right)+\Omega_k  + 3\Omega_D \omega_D   \right]. \label{52}
\end{eqnarray}
Then, the final expression of $q$ is obtained substituting the expression of the EoS parameter $\omega_D$ given in Eq. (\ref{42})  in Eq. (\ref{52}).\\
We can now derive the important quantities of our model in the limiting case corresponding to $\alpha = \beta = 0$ ($\gamma_c =1$), which means no corrections terms in the energy density. \\
The expression of the energy density $\rho_D$ given in Eq. (\ref{18}) reduces to:
\begin{eqnarray}
 \rho_D = \frac{3c^2 \phi^2}{4\omega}R, \label{54}
\end{eqnarray}
which is equivalent to the expression of the Ricci DE (RDE) in Brans-Dicke cosmology studied by Feng \cite{fengBD}, which found that this model can explain well the current acceleration of the universe. \\
From Eq. (\ref{54}), using the definitions of $R$ and $\phi$ along with Eqs. (\ref{21}) and (\ref{25}), we find the Hubble parameter $H$ reduces to:
\begin{eqnarray}
H &=& \frac{6c^2}{12c^2-1}\left(\frac{1}{t}\right).\label{55}
\end{eqnarray}
Moreover, the expression of the Ricci scalar curvature $R$ becomes:
\begin{eqnarray}
R &=& \frac{36c^2}{(12c^2-1)^2}\left(\frac{1}{t^2}\right).\label{56}
\end{eqnarray}
The EoS parameter $\omega_D$ given in Eq. (\ref{42}) reduces to:

\begin{eqnarray}
\omega_D &=& \left[ \frac{1}{27c^2}\left(1 -\frac{8\rho_{\phi}}{\phi^2R }\right) \right] \left(2n^2\omega -6n -3 \right)   \nonumber \\
&& - \frac{1}{9}\left( 4n^2\omega -6n-3 \right)\left(\frac{1+\Omega_k - \Omega_{\phi}}{\Omega_D}\right).\label{57}
\end{eqnarray}
The expression of the deceleration parameter $q$ can be obtained substituting the expression of $\omega_D$ obtained in Eq.(\ref{57}) in Eq. (\ref{52}).\\
Moreover, we can also observe that, in the limiting case of the Einstein's gravity (i.e., $\phi^2=4\omega M_p^2$, $n=0$ and $\Omega_{\phi} = \rho_{\phi} = 0$), we recover the same results found in Pasqua et al. \cite{pasqua}.

\section{Conclusions}
In this paper, we studied the logarithmic entropy corrected HDE (LECHDE) withinfrared (IR) cut-off given by the average radius of Ricci scalar curvature. The logarithmic correction is motivated from the LQG, which represents one of the most promising theories of quantum gravity. We started considering a non-flat FLRW background in BD gravitational theory.
This theory involves a scalar field $\phi$ which accounts for a dynamical gravitational constant.
We assumed an ansatz by which the BD scalar field $\phi$ evolves with the expansion of the
universe. Then, we established a correspondence between the field and the R-LECHDE to
study its dynamic, which is governed by some dynamical parameters like the EoS parameter $\omega_D$, evolution of energy density parameter $\Omega'_D$ and the deceleration parameter $q$. We calculated them in the non-flat background with interaction between R-LECHDE and DM. Moreover, we calculated the limiting cases corresponding to $\alpha = \beta =0$, (i.e., $\gamma_c =1$ then no corrections into the expression of the energy density) and Einstein's gravity. In particular, in the case of no corrections, we obtained the same model studied by Feng \cite{fengBD} which is able to describe the acceleration of the universe and, in the limiting case of Einstein gravity, we found similar results with a previous work made by Pasqua et al. \cite{pasqua}).\\
The model we studied has many parameters which are not still accurately fixed. Future high precision cosmological observations, hopefully, may be able to determine the fine property of the interacting entropy-corrected
holographic model of DE in the framework of BD gravity and, then, to reveal some significant features of DE.


\begin{thebibliography}{99}

\bibitem{1} de Bernardis, P., et al., Nature, \textbf{404}, 955 (2000).
\bibitem{1a} Perlmutter, S., et al., Astrophys. J., \textbf{517}, 565 (1999).
\bibitem{1b} Riess, A. G.,  et al., Astron. J., \textbf{116}, 1009 (1998).
\bibitem{1c} Seljak, U.,  et al.\, Phys. Rev. D, \textbf{71}, 103515 (2005).

\bibitem{2} Padmanabhan, T.,  Phys. Rep., \textbf{380}, 235, (2003).
\bibitem{2a} Cai, Y. F., Saridakis, E. N., Setare, M. R., \& Xia, J. Q., Phys. Rep., \textbf{493}, 1 (2010).
\bibitem{2b} Copeland, E. J., Sami, M., \& Tsujikawa, S., International Journal of Modern Physics D, \textbf{15}, 1753 (2006).

\bibitem{copeland-2006} Copeland, E.~J., Sami, M., \& Tsujikawa, S., International Journal of Modern Physics D, \textbf{15}, 1753  (2006).

\bibitem{delcampo} del Campo, S.,  Herrera, R., Pavon, D., J. Cosmol. Astropart. Phys. \textbf{0901}, 020 (2009).
\bibitem{delcampoa} Leon, G.,  Saridakis, E. N., Phys. Lett. B \textbf{693}, 1 (2010).
\bibitem{delcampob} Jimenez, J. B., Maroto, A. L., AIP Conf. Proc. \textbf{1122}, 107 (2009).
\bibitem{delcampoc} Berger, M. S., Shojae, H., Phys. Rev. D \textbf{73}, 083528 (2006).
\bibitem{delcampod} Zhang, X., Mod. Phys. Lett. A \textbf{20}, 2575 (2005).
\bibitem{delcampoe} Griest, K., Phys. Rev. D \textbf{66}, 123501 (2002).
\bibitem{delcampof} Jamil, M., Rahaman, F., Eur. Phys. J. C \textbf{64}, 97 (2009).
\bibitem{delcampog} Jamil, M., Saridakis, E. N., Setare, M. R., Phys. Rev. D \textbf{81}, 023007 (2010a).
\bibitem{delcampoh} Jamil, M.,  Saridakis, E. N., J. Cosmol. Astropart. Phys., \textbf{07}, 028 (2010b).
\bibitem{delcampoi} Jamil, M.,  Farooq, M. U., J. Cosmol. Astropart. Phys., \textbf{03}, 001 (2010c).
\bibitem{delcampol} M. Jamil, A. Sheykhi, M. U. Farooq, Int. J. Mod. Phys. D, \textbf{19}, 1831 (2010d).

\bibitem{3} Li, M.,  Physics Letters B, \textbf{603}, 1 (2004).
\bibitem{3a} Myung, Y. S.,  Physics Letters B, \textbf{649}, 247 (2007).
\bibitem{3b} Myung, Y. S., \& Seo, M. G., Physics Letters B, \textbf{671}, 435 (2009).

\bibitem{4} Huang, Q. G., \&  Li, M., J. Cosmol. Astropart. Phys., \textbf{8}, 13 (2004).

\bibitem{5} 't Hooft, G.,  International Journal of Modern Physics D, \textbf{15}, 1587 (2006).
\bibitem{5a} Susskind, L., Journal of Mathematical Physics, \textbf{36}, 6377 (1995).
\bibitem{5b} Bigatti, D.,  \& Susskind, L.,  Strings, Branes and Gravity  TASI \textbf{99}, 883 (2001).

\bibitem{6} Fischler, W., \& Susskind, L., 1998, arXiv:hep-th/9806039

\bibitem{7} Cohen, A. G., Kaplan, D. B., \&  Nelson, A. E., Physical Review Letters, \textbf{82}, 4971 (1999)

\bibitem{8} Guberina, B., Horvat, R.,  \& Nikolic, H., J. Cosmol. Astropart. Phys., \textbf{1}, 12 (2007)

\bibitem{9} Bekenstein, J. D., Phys. Rev. D, \textbf{7}, 2333 (1973)
\bibitem{9a}  Hawking, S. W., Phys. Rev. D, \textbf{13}, 191 (1976)

\bibitem{10} Chen, B., Li, M., \& Wang, Y., Nuclear Physics B, \textbf{774}, 256 (2007).

\bibitem{11} Jamil, M.,  Saridakis, E. N., \& Setare, M. R., Physics Letters B, \textbf{679}, 172 (2009).

\bibitem{12} Xu, L., J. Cosmol. Astropart. Phys., \textbf{9}, 16 (2009).
\bibitem{12a} Sadjadi, H. M., \& Jamil, M., General Relativity and Gravitation, \textbf{43}, 1759 (2011).
\bibitem{12b} Karami, K.,  Jamil, M.,  Roos, M., Ghaffari, S.,  \& Abdolmaleki, A., Astrophys. Space Sci., \textbf{340}, 175 (2012).
\bibitem{12c} Karami, K., \& Fehri, J., International Journal of Theoretical Physics, \textbf{49}, 1118 (2010).
\bibitem{12d} Jamil, M., \& Farooq, M. U., International Journal of Theoretical Physics, \textbf{49}, 42 (2010).
\bibitem{12e} Jamil, M., Farooq, M. U., \& Rashid, M. A., European Physical Journal C, \textbf{61}, 471 (2009).
\bibitem{12f} Jamil, M.,  Karami, K., \& Sheykhi, A., International Journal of Theoretical Physics, \textbf{50}, 3069 (2011).
\bibitem{12g} Jamil, M., \& Sheykhi, A., International Journal of Theoretical Physics, \textbf{50}, 625 (2011).

\bibitem{13} Wang, B., Gong Y., \& Abdalla, E., Physics Letters B, \textbf{624}, 141 (2005).
\bibitem{13a} Wang, B., Lin, C. Y., \& Abdalla, E., Physics Letters B, \textbf{637}, 357 (2006).
\bibitem{13b} Wang, B.,  Lin, C. Y.,  Pav{\'o}n, D., \& Abdalla, E., Physics Letters B, \textbf{662}, 1 (2008).
\bibitem{13c} Sheykhi, A., Classical and Quantum Gravity, \textbf{27}, 025007 (2010).

\bibitem{14} Chattopadhyay, S., \& Debnath, U., Astrophys. Space Sci., \textbf{319}, 183 (2009).
\bibitem{14a} Karami, K., \& Fehri, J., Physics Letters B, \textbf{684}, 61 (2010).
\bibitem{14b} Karami, K.,  Khaledian, M. S., \& Jamil, M., Phys. Scr., \textbf{83}, 025901 (2011).

\bibitem{15} Fen, C. J., \& Li, X. Z., Physics Letters B, \textbf{679}, 151 (2009).
\bibitem{15a} Feng, C. J., \& Zhang, X., Physics Letters B, \textbf{680}, 399 (2009).
\bibitem{15b} Wei, H., Nuclear Physics B, \textbf{819}, 210 (2009).
\bibitem{15c} Bisabr, Y., General Relativity and Gravitation, \textbf{41}, 305 (2009).
\bibitem{15d} Nozari, K., \& Rashidi, N., International Journal of Modern Physics D, \textbf{19}, 219 (2010).
\bibitem{15e} Nozari, K., \& Rashidi, N., International Journal of Theoretical Physics, \textbf{48}, 2800 (2009).
\bibitem{15f} Jamil, M., Sheykhi, A., \& Farooq, M. U., International Journal of Modern Physics D, \textbf{19}, 1831 (2010).
\bibitem{15g} Karami, K., \& Khaledian, M. S., Journal of High Energy Physics, \textbf{3}, 86 (2011).
\bibitem{15h} Setare, M. R., Physics Letters B, \textbf{644}, 99 (2007).
\bibitem{15i} Setare, M. R., \& M. Jamil, M.,  EPL (Europhysics Letters), \textbf{92}, 49003 (2010).
\bibitem{15l} Setare, M. R., \& M. Jamil, M., Physics Letters B, \textbf{690}, 1 (2010).

\bibitem{16} Feng, C., Wang, B., Gong, Y., \& Su, R. K., J. Cosmol. Astropart. Phys., \textbf{9}, 5 (2007).
\bibitem{16a} Wang, B., Zang, J.,  Lin, C. Y., Abdalla, E., \& Micheletti, S., Nuclear Physics B, \textbf{778}, 69 (2007).
\bibitem{16b} Wu, Q., Gong, Y., Wang, A., \& Alcaniz, J. S., Physics Letters B, \textbf{659}, 34 (2008).
\bibitem{16c} Li, M., Li, X. D., Wang, S., Wang, Y., \& Zhang, X., J. Cosmol. Astropart. Phys., \textbf{12}, 14 (2009).
\bibitem{16d} Lu, J., Saridakis, E. N., Setare, M. R., \& Xu, L., J. Cosmol. Astropart. Phys., \textbf{3}, 31 (2010).
\bibitem{16e} Zhang, X., Physics Letters B, \textbf{683}, 81 (2010).

\bibitem{17} Banerjee, R., Gangopadhyay, S., \& Modak, S. K., Physics Letters B, \textbf{686}, 181 (2010).
\bibitem{17a} Banerjee, R., \& Modak, S. K., Journal of High Energy Physics, \textbf{5}, 63 (2009).
\bibitem{17b} Modak, S. K., Physics Letters B, \textbf{671}, 167 (2009).

\bibitem{18} Banerjee, R., \& Majhi, B. R., Physics Letters B, \textbf{662}, 62 (2008).
\bibitem{18a} Banerjee, R.,  \& Ranjan Majhi, B., Journal of High Energy Physics, \textbf{6}, 95 (2008).
\bibitem{18b} Zhang, J., 2008, Physics Letters B, \textbf{668}, 353 (2008).
\bibitem{18c}  Karami, K., \& Abdolmaleki, A., Phys. Scr., \textbf{81}, 055901 (2010). 
\bibitem{18d} Karami, K.,  Sheykhi, A.,  Jamil, M.,  Azarmi, Z.,  \& Soltanzadeh, M. M., General Relativity and Gravitation, \textbf{43}, 27 (2011).

\bibitem{19} Ashtekar, A., Baez, J., Corichi, A., \& Krasnov, K., Physical Review Letters, \textbf{80}, 904 (1998).
\bibitem{19a} Rovelli, C., Physical Review Letters, \textbf{77}, 3288 (1996).
\bibitem{19b} Ghosh, A., \& Mitra, P., Phys. Rev. D, \textbf{71}, 027502 (2005).
\bibitem{19c} Medved, A. J., \& Vagenas, E. C., Phys. Rev. D, \textbf{70}, 124021 (2004).
\bibitem{19d} Meissner, K. A., Classical and Quantum Gravity, \textbf{21}, 5245 (2004).

\bibitem{20} Wald, R. M., Chicago, University of Chicago Press, 1984, 504 p.

\bibitem{21} Cai, Y. F., Liu, J., \& Li, H., Physics Letters B, \textbf{690}, 213 (2010).

\bibitem{22} Wei, H., Communications in Theoretical Physics, \textbf{52}, 743 (2009).

\bibitem{23} Gong, Y., Phys. Rev. D, \textbf{70}, 064029 (2004).
\bibitem{23a} Kim, H.,  Lee, H. W., \& Myung, Y. S., Physics Letters B, \textbf{628}, 11 (2005).

\bibitem{24} Kim, H.,  Lee, H. W., \& Myung, Y. S., 2005, arXiv:hep-th/0501118.
\bibitem{24a} Gong, Y., Phys. Rev. D, \textbf{61}, 043505 (2000).
\bibitem{24b} Nayak, B., \& Singh, L. P., Modern Physics Letters A, \textbf{24}, 1785 (2009).
\bibitem{24c} Xu, L., Li, W., \& Lu, J., European Physical Journal C, \textbf{60}, 135 (2009).
\bibitem{24d} Sheykhi, A., Phys. Rev. D, \textbf{81}, 023525 (2010).
\bibitem{24e} Sheykhi, A., \& Jamil, M., Physics Letters B, \textbf{694}, 284 (2011).

\bibitem{25} Sheykhi, A., Physics Letters B, \textbf{681}, 205 (2009).

\bibitem{fengBD} Feng, C.-J.\ 2008, arXiv:0806.0673 .

\bibitem{gao} Gao, C.,  Wu, F., Chen, X., \& Shen, Y. G.,  Phys. Rev. D, \textbf{79}, 043511 (2009).

\bibitem{cai} Cai, R. G., Hu, B., \& Zhang, Y., Communications in Theoretical Physics, \textbf{51}, 954 (2009).

\bibitem{mio2} Pasqua, A., Jamil, M., Myrzakulok, R., \& Majeed, B., Phys. Scr., \textbf{86}, 045004 (2012).

\bibitem{feng3} Feng, C. J., Physics Letters B, \textbf{670}, 231 (2008).

\bibitem{feng1}  Feng, C. J., Physics Letters B, \textbf{676}, 168 (2009).

\bibitem{feng2} Feng, C. J., Physics Letters B, \textbf{672}, 94 (2009).

\bibitem{feng4}  Feng, C. J., \& Li, X. Z., Physics Letters B, \textbf{680}, 355 (2009).

\bibitem{xu} Xu, L., \& Wang, Y., J. Cosmol. Astropart. Phys., \textbf{6}, 2 (2010).

\bibitem{26} Arik, M., {\c C}alik, M. C., \& Sheftel, M. B., International Journal of Modern Physics D, \textbf{17}, 225  (2008).
\bibitem{26-1} Arik, M., \& {\c C}alik, M. C., Modern Physics Letters A, \textbf{21}, 1241 (2006).

\bibitem{28} Banerjee, N., \& Pav{\'o}n, D., Physics Letters B, \textbf{647}, 477 (2007).

\bibitem{29} Xu, L.,  Lu, J., \& Li, W.,\ 2009, arXiv:0905.4174.

\bibitem{30} Acquaviva, V., \& Verde, L., J. Cosmol. Astropart. Phys., \textbf{12}, 1 (2007).
\bibitem{30-1} Bertotti, B.,  Iess, L., \& Tortora, P., Nature, \textbf{425}, 374  (2003).

\bibitem{sceichi} Sheykhi, A., Physics  Letters B, \textbf{681}, 205 (2009).

\bibitem{q1} Amendola, L., \& Tocchini-Valentini, D., Phys. Rev. D \textbf{64}, 043509 (2001).
\bibitem{q1-1} Amendola, L., \& Tocchini-Valentini, D., Phys. Rev. D \textbf{66}, 043528 (2002).
\bibitem{q1-2} Jamil, M., Saridakis, E. N., \& Setare, E. R., Phys. Lett. B \textbf{679}, 172 (2010).
\bibitem{q1-3} Setare, M. R., \& Jamil, M., Phys. Lett. B \textbf{690}, 1 (2010).
\bibitem{q1-4} Sheykhi, A., \& Jamil, M., Phys. Lett. B \textbf{694}, 284 (2011).
\bibitem{q1-5} Farooq, M. U., Jamil, M., \& Rashid, M. A., Int. J. Theor. Phys \textbf{49}, 2278 (2010).
\bibitem{q1-6} Jamil, M., \& M. Farooq, M. U., Int. J. Theor. Phys. \textbf{49}, 42 (2010).
\bibitem{q1-7} Zimdahl, W., \& Pavon, D., Phys. Lett. B \textbf{521}, 133 (2001).
\bibitem{q1-8} Zimdahl, W., \&  Pavon, D., Gen. Rel. Grav. \textbf{35}, 413 (2003).
\bibitem{q1-9} Setare, M. R., \& Jamil, M., J. Cosmol. Astropart. Phys. \textbf{02}, 010 (2010).
\bibitem{q1-10} Jamil. M., \&  Farooq, M. U., J. Cosmol. Astropart. Phys. \textbf{03}, 001 (2010).

\bibitem{q2} Bertolami, O., Gil Pedro, F., \& Le Delliou, M., Physics Letters B, \textbf{654}, 165 (2007).
\bibitem{q2-2} Jamil, M., \& Rashid, M.A.,  European Physical Journal C, \textbf{58}, 111 (2008).

\bibitem{q3} Feng, C., Wang, B., Gong, Y., \& Su, R. K., J. Cosmol. Astropart. Phys., \textbf{9}, 5 (2007).

\bibitem{q4} Ichiki, K., et. al., J. Cosmol. Astropart. Phys. \textbf{06}, 005 (2008).

\bibitem{pasqua} Pasqua, A.,  Khodam-Mohammadi, A.,  Jami, M., \& Myrzakulov, R., Astrophys. Space Sci. \textbf{340}, 199 (2012).

\end{thebibliography}
\end{document}